\documentclass[12pt,preprint]{aastex} 
 
\shorttitle{Depletion and deuterium fractionation in pre-stellar cores} 
\shortauthors{Bacmann et al.} 
 
\begin{document} 
 
 
\title{CO depletion and deuterium fractionation in pre-stellar cores} 
 
 
\author{A. Bacmann\altaffilmark{1,2}, B. Lefloch\altaffilmark{3},
C. Ceccarelli\altaffilmark{3,4}, J. Steinacker\altaffilmark{2},
A. Castets\altaffilmark{4}, L. Loinard\altaffilmark{5}} 

 
\altaffiltext{1}{European Southern Observatory, Karl-Schwarzschild-Str. 2, D-85748 Garching-bei-M\"unchen, Germany; abacmann@eso.org}
\altaffiltext{2}{Astrophysikalisches Institut und Universit\"ats-Sternwarte, Schillerg\"a{\ss}chen 2-3, D-07745 Jena} 
\altaffiltext{3}{Laboratoire d'Astrophysique, Observatoire de Grenoble, BP\,53, F-38041 Grenoble C\'edex 9, France} 
\altaffiltext{4}{Observatoire de Bordeaux, 2 rue de l'Observatoire, BP\,89, F-33270 Floirac, France} 
\altaffiltext{5}{Instituto de Astronom\'{\i}a, UNAM, Apartado Postal 72-3 (Xangari), 58089 Morelia, Michoac\'an, Mexico}

 
\begin{abstract} 
   We report the detection of D$_2$CO in a sample of starless dense cores,  
in which we previously measured the degree of CO depletion. The deuterium  
fractionation is found extremely high, ${\rm [D_2CO]/[H_2CO]}\sim 1-10$\%, 
similar to that reported in low-mass protostars. This provides convincing  
evidence that D$_2$CO is formed in the cold pre-stellar cores, and later  
desorbed when the gas warms up in protostars. We find that the cores with the  
highest CO depletions have also the largest [D$_2$CO]/[H$_2$CO] ratios,  
supporting the theoretical prediction that deuteration increases with  
increasing CO depletion.  
\end{abstract} 
 
 
\section{Introduction} 
 
Multiply deuterated molecules have drawn much attention in the past few years, 
with the detections of D$_2$CO \citep{turner90}, ND$_2$H  
\citep{rtc00} and even ND$_3$ \citep{lrg02,vdt02}.  
D$_2$CO was found to be particularly 
abundant (with [D$_2$CO]/[H$_2$CO]$\sim 5 - 40$\,\%) 
in low-mass protostellar environments \citep{ccl98,lcc02}.  
The mechanisms leading to such high degrees of 
deuteration remain however debated. According to \citet{t83} and  
\citet{ctr97}, high deuterium fractionation can be obtained via active  
grain chemistry: the deuteration of the molecules, stuck onto the grains, 
 is regulated by the {\it atomic} D/H ratio of the gas accreting into the mantles. 
The other major way to  
forming deuterated molecules is through pure gas-phase reactions in presence 
of strong CO (and other molecules) depletion \citep[hereafter RMa]{rm00a}. In this case,  
the deuterium is propagated to other molecules mostly 
by reactions with H$_2$D$^+$, and the [H$_2$D$^+$]/[H$_3^+$] ratio, which is 
also proportional to the atomic D/H ratio, plays a crucial 
role in determining the molecular deuterium ratio. 
In both approaches, a low temperature ($T\le 20$\,K) is necessary 
to obtain enhanced D/H or [H$_2$D$^+$]/[H$_3^+$] ratios and therefore enhanced 
levels of molecular deuteration.  
 
The gas-grain interaction mechanism was  
favoured to interpret the observations towards IRAS16293-2422 \citep[and references therein]{clc01}. In this scheme, D atoms react with the CO  
molecules on the grain surface {\it during the colder, pre-collapse stage}, and the  
obtained molecules are later released into  
the gas phase through heating from the newly-formed protostars. Pure gas phase models 
\citep[hereafter RMb]{rm00b} could however well account for the observed 
[ND$_2$H]/[NH$_3$] ratios found by \citet{rtc00}. Recently, D$_2$CO 
was detected in L1689N, the cloud which IRAS16293 is embedded in, with an 
abundance ratio similar to that found at the border of the envelope of the protostar 
\citep{cvt02}. This detection posed new problems, since none of 
the aforementioned models can account for the abundance of D$_2$CO in that cloud. 
In order to have a high atomic D/H ratio, both models 
require strong CO and O depletion, which is not consistent with the high atomic O 
abundance observed \citep{ccc99,vcc00,lkp01}. Moreover, a former colder, denser  
(and possibly more depleted)  
phase cannot be invoked for L1689N. 
  
Pre-stellar cores, as (starless) progenitors of low-mass protostars, offer the 
opportunity to study the cold, dense evolutionary stage in which the deuteration of 
molecules supposedly occurred. In addition, these objects are known to show strong  
degrees of CO (among other molecules) depletion, which is a necessary requisite for 
the models: the amount of deuteration should be linked to the degree of CO 
depletion. We recently studied the degree of CO depletion in a sample of pre-stellar 
cores, and showed that the depletion degree increases with density \citep{blc02}.  
In this Letter, we report the  
successful search for D$_2$CO in those pre-stellar cores and test quantitatively 
the hypothesis that stronger CO depletions lead to larger [D$_2$CO]/[H$_2$CO] ratios.  
We first briefly describe our multi-transition H$_2$CO and D$_2$CO observations 
in Section\,\ref{observ} and present our results on the H$_2$CO abundance and D$_2$CO 
over H$_2$CO abundance ratio in Section\,\ref{results}. We then discuss our results 
and their implications on deuteration mechanisms  
and finally conclude in Section\,\ref{concl}. 
 
\section{Observations} 
 
\label{observ} 
 
We observed in August 2001 a sample of cores for which we had already discussed the CO 
depletion \citep{blc02} in various transitions of H$_2$CO and D$_2$CO 
(cf. JPL Catalog) at the IRAM 30\,m telescope on Pico Veleta in 
Spain. The spectra were taken in the 
position switching mode at the core central position (maximum of the continuum 
emission). Additional observations  (small maps around the peak)  
to check the extent of the emission of H$_2$CO(2$_{12}$-1$_{11}$) and 
H$_2$CO(3$_{13}$-2$_{12}$) were carried out  
in November 2001. The {\it rms} reached were between 20 and 80\,mK on the H$_2$CO transitions 
and between 15 and 50\,mK on the lower energy D$_2$CO transitions  
D$_2$CO(2$_{12}$-1$_{11}$) and D$_2$CO(3$_{13}$-2$_{12}$).  
In order to estimate the opacities of the H$_2$CO transitions, we observed  
in August 2002 the (2$_{12}$-1$_{11}$) transition of H$_2^{13}$CO in all  
cores but L1544.  
In all these observing runs, 
we used as backend an autocorrelator 
with spectral resolution of 40\,kHz at $100-150$\,GHz, and 80\,kHz at $200-245$\,GHz, 
yielding resolutions of $\sim 0.08-0.11$\,km\,s$^{-1}$ depending on line frequency.  
The beam and forward efficiencies at the different line frequencies can be found on 
the IRAM Web page at {\tt http://www.iram.es}. Pointing was checked every 1.5 hours and found to 
be better than $\simeq$\,3\arcsec. In the following, line fluxes are given in 
units of main beam brightness temperature. In addition to the {\it rms}  
noise, a 
calibration uncertainty of 15\% at 150\,GHz and of 20\% at 230 GHz was taken  
into account. 

\section{Results} 
 
\label{results} 
 
\begin{figure} 
\plotone{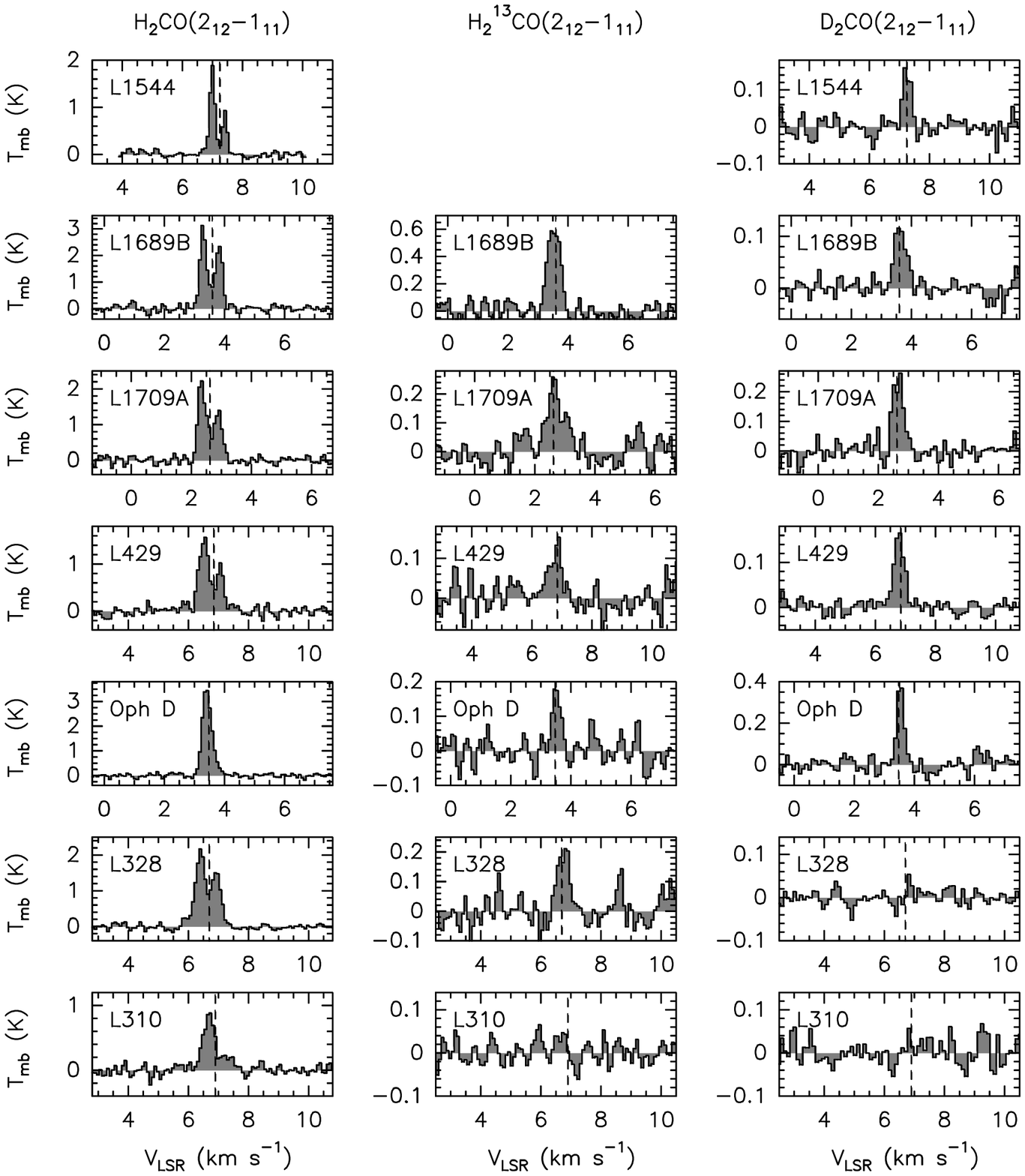} 
\caption{H$_2$CO$(2_{12}-1_{11})$, H$_2^{13}$CO$(2_{12}-1_{11})$ and  
     D$_2$CO$(2_{12}-1_{11})$ lines in the pre-stellar cores of the sample.  
     The D$_2$CO line is detected with a good signal-to-noise ratio in L1544,  
     L1689B, L1709A, L429 and Oph\,D. H$_2$CO is double-peaked in 6 of the  
     cores. The {\it lsr} velocity is shown as dashed lines.\label{lines}} 
\end{figure}

In 5 out of the 7 cores in our sample, D$_2$CO was clearly detected in at  
least one transition (Table\,\ref{obs}) 
\footnote{The non-detection of D$_2$CO($2_{21}-1_{11}$) in L328 is questionable,  
  since a 3$\sigma$ line was seen at the right velocity after 20 minutes of  
  integration on both 3\,mm receivers. Further integration did not confirm this  
  detection.} 
. The transition 
 D$_2$CO$(2_{12}-1_{11})$ is a $5\sigma-12\sigma$ detection in those 5 cores  
(Fig\,\ref{lines}). The line fluxes given in this Table were determined by integrating 
the line signal between two given velocities (typically, where the signal is larger than 
the noise). For non-detections, the upper limit was taken as 3 times the {\it rms} on 
the velocity-integrated intensity, given a linewidth close to that of detected  
lines of the same molecule.

To derive the D$_2$CO and H$_2$CO 
column densities and abundances, we used rotational diagrams (see \citealt{gl99} for a  
full description of the method).  
Rotational diagrams give a correct estimate of the column density 
only when the lines are thermally populated and optically thin. 
If, however, the lines are sub-thermally excited, the rotational column  
density can be substantially different from the real one.  
For this reason, in this article 
we focus only on the relative [D$_2$CO]/[H$_2$CO] ratio, as it is not  
affected by such uncertainty thanks to the similarity of the two molecules. 
Line opacity can also affect the rotation temperature, but has little 
effect on the column density ratio.  
This method implies identical excitation conditions in both lines. 
We estimated separately 
the ortho-H$_2$CO and para-H$_2$CO column densities, and assumed an  
ortho-to-para  
ratio of 2 for D$_2$CO. A single rotational temperature was derived  
simultaneously 
for both the ortho- and the para- H$_2$CO and then used to determine the  
D$_2$CO column density. In the cases where more than one D$_2$CO transition  
was  
detected, the rotational temperature was derived simultaneously for  
ortho-H$_2$CO,  
para-H$_2$CO and D$_2$CO as well. For L1544, the rotational temperature 
was assumed to be 7.25$\pm$2.5\,K, which is an average of the values found  
for the  
other cores, and consistent with the gas temperature determined by Bacmann et  
al. (2002). From the small maps we made around the central position,  
the H$_2$CO(3$_{13}$-2$_{12}$) emission was  
found to be extended in the cores.  
To correct for the difference in beam filling factors at 145 and 220 GHz, 
the H$_2$CO(3$_{13}$-2$_{12}$) was 
degraded to the spatial resolution of the  H$_2$CO(2$_{12}$-1$_{11}$) line.  
The intensity ratio between the original and the degraded spectra  
(close to one)  
was applied to estimate the degraded H$_2$CO(3$_{03}$-2$_{02}$)  
spectra. Note that both 
transitions are very close in frequencies and upper level energy. 
Finally, we corrected for the line opacity effects estimated by  
our observations of the H$_2^{13}$CO (2$_{12}$-1$_{11}$) transition, which suggest 
opacities of 2--12 (most of them around 3--5).  
The final H$_2$CO column densities of Table \ref{coldens} have been obtained  
by taking a $^{12}{\rm C}/^{13}{\rm C}$ isotopomer abundance ratio of 60 
(e.g. \citealt{beg00}). 
 
 

 
Incidently, the observed H$_2$CO line profiles are consistent (except for Oph D) 
with infalling gas, the lines being self-absorbed and the blue-shifted peak brighter  
than the red-shifted one \citep{lb77}. A more detailed analysis of the  
observed kinematics  
as well as the absolute abundances and ortho-to-para ratios 
will be presented in a forthcoming paper. 
The minimum and maximum values of the column densities and the  
[D$_2$CO]/[H$_2$CO]  
abundance ratios were determined by using a Monte-Carlo method to explore the  
whole range of possible values for the rotational temperature  
and integrated line flux within  
the error bars. The column densities in H$_2$CO, D$_2$CO, H$_2$, 
the D$_2$CO abundances and the [D$_2$CO]/[H$_2$CO] abundance ratios are  
given in Table\,\ref{coldens}.

D$_2$CO is found extremely abundant in all cores where it was detected. The  
[D$_2$CO]/[H$_2$CO] abundance ratios given in Table\,\ref{coldens} (between  
$\sim$\,1\% and 10\%) are comparable to the values quoted by \citet{lcc02}  
for low-luminosity protostars. The H$_2$CO ortho/para ratio  
was found around 2.5-3 (except for L310 where it was $\sim 1.8$), 
which is consistent with a gas-phase formation of H$_2$CO  
(e.g. \citealt{di99}).
 
\section{Discussion \& Conclusions: Relation with CO depletion} 
 
\label{discus} 
 
\begin{figure} 
\plotone{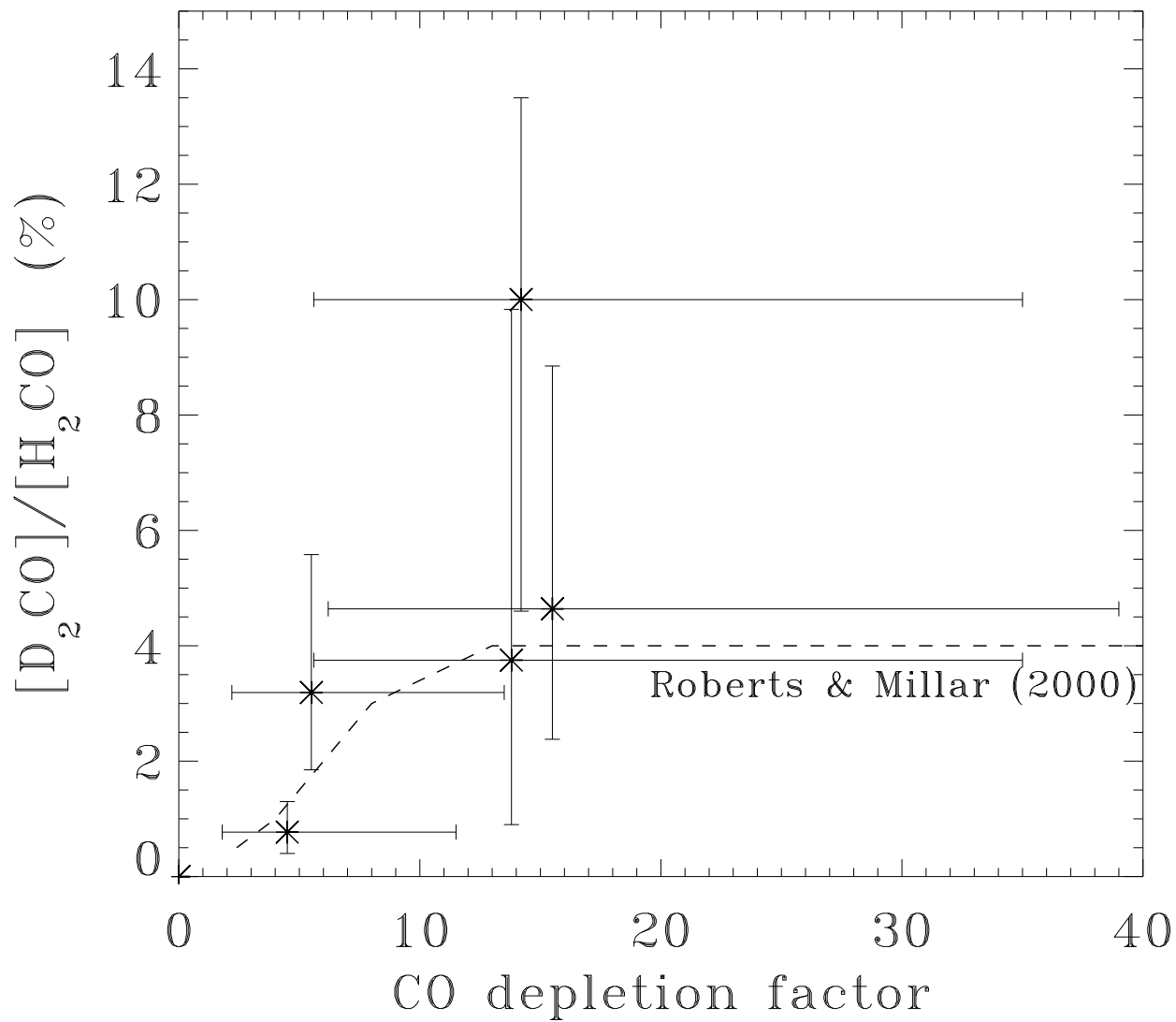} 
\caption{Comparison of [D$_2$CO]/[H$_2$CO] abundance ratio and CO depletion 
     factor (from \citet{blc02} for the dense cores in the sample.  
     The [D$_2$CO]/[H$_2$CO] vs. CO depletion relation derived from the model 
     of RMb is represented by the dashed curve.\label{d2co}} 
\end{figure}

The high abundance of D$_2$CO in pre-stellar cores provides convincing  
evidence  
that the D$_2$CO seen in low-luminosity protostars originates in the cold,  
pre-stellar stage, as was suggested by \citet{ccl98} and their 
subsequent works (see introduction). 
In the gas-phase schemes, high deuterium fractionations of a few percent  
(up to $\sim 40\%$) like those measured around young low-luminosity protostars  
\citep{lcc02} can only be achieved when the gas phase is highly  
depleted of molecules and atoms reacting with H$_3^+$ (RMa, RMb).  
It is also the case in gas-grain chemistry models, 
for which high deuterium fractionations are obtained when the atomic D/H is high, 
which is in turn favoured when the gas phase is depleted of CO.  
In order to test this hypothesis, we plotted in Fig.\,\ref{d2co} the 
[D$_2$CO]/[H$_2$CO] ratio against the CO depletion degree estimated in \citet{blc02}  
- and their Table 2.  As can be seen on the plot, the scatter 
and error bars are large (especially for L1544 for which the data are not as  
complete), although cores with high depletion factors tend to 
have high deuteration fractions in formaldehyde.  In this figure, the  
error bars on the CO depletion can be mostly attributed to the uncertainty  
on the dust extinction coefficient $\kappa_{1.3{\rm mm}}$ at 1.3\,mm (a factor of  
$\sim 2$). Despite the large scatter, the trend that we observe is consistent 
with an increase of deuteration with the CO depletion.  
To give a quantitative estimate, the Pearson correlation coefficient is 0.7, 
indicating a large probability that the two quantities, [D$_2$CO]/[H$_2$CO]  
and CO depletion, are indeed correlated. 
The models also state that the gas phase should be depleted of O, which 
is an important H$_3^+$ destroyer as well, only a small factor 2.5 less  
efficient than CO, 
according to the UMIST database (http:www.rate99.co.uk/).  
We have for the time being no {\it direct} information on the amount of  
atomic oxygen 
still in the gas phase in these cores,  
but some indirect indications that O is not much depleted indeed.  
On the one hand, large quantities of gaseous atomic oxygen have been 
observed in  
several cold, CO depleted, molecular clouds \citep{ccc99,vcc00,lgp02}. 
On the other hand, a detailed modeling of the L1544 core (also in our  
sample) molecular  
composition also suggests a large atomic oxygen abundance in the gas phase  
\citep{cas02}. 
The apparent correlation between [D$_2$CO]/[H$_2$CO] and CO depletion 
and the very high fractionation observed may suggest 
either a similar degree of oxygen depletion with respect to CO 
(which would disagree with the aforementioned arguments)  
or a lower role of O in the fractionation 
(yet not supported by theoretical considerations). 
In the last case, the relatively large scatter observed in Fig. \ref{d2co}  
may be due to  
the presence of O in the gas phase. 
 
The question remains whether D$_2$CO is formed in the gas phase \citep{rm00b}  
and frozen out onto the grains or directly formed on  
the grains as suggested by \citet{t83}. If D$_2$CO is formed on the  
grains, there must exist an efficient desorption mechanism to release part 
of the D$_2$CO into the gas phase, even in such cold ($T\sim 10$\,K) environments  
as pre-stellar cores. The depletion timescale ($\sim 10000$\,yr - to be compared 
with estimated lifetimes of 10$^6$\,yr, see \citealt{bme86}) derived in  
\citet{blc02} suggests that desorption mechanisms indeed exist, 
since it is highly unlikely that all cores from the sample have only reached 1\% of  
their lifetime. The gas-grain model is thus not ruled out by observations 
of D$_2$CO in the gas phase, even in objects with $T\sim 10$\,K.  
For the rest, the same remarks as on the cloud L1689N  
\citep{cvt02} apply here.  
In the absence of O depletion, the atomic D/H ratio is predicted to be around 0.01  
(RMa), which is not sufficient to account for the high levels of  
deuteration observed here (to account for 
a $\rm [D_2CO]/[H_2CO]$ ratio of 0.05, the atomic D/H ratio would have to be as high  
as 0.3 - \citealt{css02}). The dashed curve in Fig.\,\ref{d2co} 
represents the predictions of the gas-phase RMb model, 
which assumes relatively large O depletion as well.  
Our data compare rather well with those predictions, with the exception 
of Oph D, which shows a larger  [D$_2$CO]/[H$_2$CO] ($\sim$\,10\%) than predicted. 
However, we  point out that another key parameter in the Roberts \& Millar  
model is the electron fractional abundance. 
The predictions reported in Fig. \ref{d2co} have been taken from the 
Fig. 3 of RMb, who assumed an 
electron fractional abundance around $10^{-7}$. This is much larger than the value 
of $X(\rm{e})\sim 10^{-9}$ actually found by \citet{cwz02} in L1544. 
Fig. 5 of RMa suggests that larger [D$_2$CO]/[H$_2$CO] 
ratios could be reached if the ionisation fraction $X(\rm{e})$ is lowered. 
It is also worth mentioning that the absolute D$_2$CO abundances we measured   
for the degrees of CO depletion found in the cores (Table\,\ref{coldens})  
are generally smaller than those  
calculated by RMb (between 10$^{-10}$ for low depletion  
factors and 1.6$\times 10^{-11}$ for depletion factors of 16), though in the end,  
we find deuterium fractionations similar or higher than those predicted by the model.  
This discrepancy is strongly marked (one order of magnitude) for the cores with  
little depletion (L1689B, L1709A and L328).  
To summarize, gas-phase models, though they can account for the  relative deuterium 
fractionation in most of our objects, cannot explain all of our data.  
Key points that still need to be fully understood are the depletion 
of atomic oxygen and its role, as well as the ionization fraction in pre-stellar cores. 
 
 
\label{concl} 
 
\acknowledgments 
 
      A.B. acknowledges support by the German \emph{Deut\-sche For\-schungs\-ge\-mein\-schaft,  
      DFG\/} in the course of this work. We thank the IRAM staff for their help during the observations 
      and the Time Allocation Committee for awarding of the telescope time.

\clearpage  
 
\clearpage 
 
\begin{deluxetable}{cccccccc} 
\tablewidth{0pt} 
\tablehead{ 
\colhead{Core} & \colhead{L1544} & \colhead{L1689B} & \colhead{L1709A} & \colhead{L310} & \colhead{L328} & \colhead{L429} & \colhead{Oph D}} 
\tabletypesize{\scriptsize} 
\tablecaption{Integrated areas of lines observed (K km s$^{-1}$) for the 7 cores of the sample. Upper limits are given for $3\sigma$ (see text).  A dash means that no observation was performed.\label{obs}} 
\startdata 
H$_2$CO($2_{12}-1_{11}$) & $0.62\pm 0.11$   & $1.63\pm 0.28$   & $1.24\pm 0.22$   & $0.40\pm 0.08$ & $1.52\pm 0.25$ & $0.93\pm 0.17$   & $1.43\pm 0.24$ \\  
H$_2$CO($2_{02}-1_{01}$) & $-$   & $1.69\pm 0.30$   & $1.16\pm 0.20$   & $0.45\pm 0.11$ & $1.23\pm 0.22$ & $0.66\pm 0.13$   & $1.40\pm 0.24$ \\  
H$_2$CO($3_{03}-2_{02}$) & $0.26\pm 0.07$   & $1.04\pm 0.28$   & $0.59\pm 0.14$   & $0.31\pm 0.12$ & $0.34\pm 0.10$ & $0.31\pm 0.08$   & $0.58\pm 0.13$ \\ 
H$_2$CO($3_{22}-2_{21}$) & $<0.083$         & $<0.18$          & $<0.068$         & $<0.237$       & $<0.090$       & $<0.054$         & $<0.076$ \\ 
H$_2$CO($3_{13}-2_{12}$) &  $-$             & $1.46\pm 0.37$   & $1.10\pm 0.27$   & $0.30\pm 0.09$ & $0.86\pm 0.21$ & $0.61\pm 0.16$   & $0.96\pm 0.23$ \\ 
D$_2$CO($2_{12}-1_{11}$) & $0.052\pm 0.012$ & $0.058\pm 0.012$ & $0.11\pm 0.02$   & $<0.046$       & $<0.020$       & $0.071\pm 0.012$ & $0.12\pm 0.02$ \\ 
D$_2$CO($3_{13}-2_{12}$) &  $-$             & $<0.14$          & $0.089\pm 0.032$ & $<0.18$        & $-$            & $0.029\pm 0.013$ & $0.14\pm 0.04$ \\ 
D$_2$CO($4_{13}-3_{12}$) &  $-$             & $<0.24$          & $-$              & $<0.16$        & $<0.25$        & $<0.061$         & $0.069\pm 0.033$ \\ 
D$_2$CO($4_{04}-3_{03}$) & $<0.05$          & $0.11\pm 0.05$   & $<0.094$         & $<0.27$        & $<0.065$       & $<0.043$         & $<0.22$ \\ 
H$_2^{13}$CO($2_{12}-1_{11}$) & $-$   & $0.33\pm 0.02$   & $0.13\pm 0.02$   & $<0.04$ & $0.10\pm 0.01$ & $0.07\pm 0.01$   & $0.04\pm 0.01$ \\  
\enddata 
\end{deluxetable}

\clearpage 
 
\begin{deluxetable}{lcccccc} 
\tablecaption{Rotational temperatures, H$_2$CO and D$_2$CO column densities observed in a 17$\arcsec$ beam (beamsize at the frequency of the $\rm H_2CO(2_{12}-1_{11})$ transition), as well as [D$_2$CO]/[H$_2$CO] abundance ratios, H$_2$ column densities (from Bacmann et al. 2002) and D$_2$CO abundances $X_{\rm D_2CO}$ ($X_{\rm D_2CO}=N_{\rm D_2CO}/N_{\rm H_2}$). The H$_2$CO column density values are corrected for opacity. Upper limits are $3\sigma$. Figures follow the convention $a(b)=a\times 10^{b}$. \label{coldens} 
} 
\tablecolumns{7} 
\tablehead{\colhead{Core} & \colhead{$T_{\rm rot}$} & \colhead{$N_{\rm H_2CO}$} & \colhead{$N_{\rm D_2CO}$} & \colhead{$\rm [D_2CO]/[H_2CO]$} & \colhead{$N_{\rm H_2}$} & \colhead{$X_{\rm D_2CO}$} \\ \colhead{} & \colhead{(K)} & \colhead{(cm$^{-2}$)} & \colhead{(cm$^{-2}$)} & \colhead{$(\%)$} & \colhead{(cm$^{-2}$)} & \colhead{}} 
\startdata 
L1544   & 7.3$^{+2.5}_{-2.5}$ & 2.5$^{+8.1}_{-1.4}$ (13) & 9.0$^{+5.3}_{-2.2}$ (11) & 4$^{+6}_{-3}$ & 1.6 (23) & $5.6 (-12)$ \\ 
L1689B  & 8.2$^{+1.1}_{-1.9}$ & 1.3$^{+0.9}_{-0.5}$ (14) & 9.9$^{+3}_{-2}$ (11) & 0.8$^{+0.5}_{-0.4}$ & 1.4 (23) & $7.1 (-12)$ \\ 
L1709A  & 7.2$^{+2.0}_{-0.9}$ & 5.6$^{+3.7}_{-2.3}$ (13) & 1.8$^{+0.6}_{-0.3}$ (12) & 3$^{+2.5}_{-1}$ & 1.4 (23)  & $1.3 (-11)$\\ 
L310    & 6.9$^{+1.9}_{-1.8}$  & 3.5$^{+12.3}_{-0.8}$ (12) & $<8.6$ (11) & $<23$ & 7.3 (22) & $< 1.2 (-11)$ \\ 
L328    & 5.9$^{+0.2}_{-0.2}$  & 4.3$^{+1.8}_{-1.4}$ (13) & $<5.8$ (11) & $<2$ & 9.5 (22) & $< 6.1 (-12)$ \\ 
L429    & 6.6$^{+0.9}_{-1.2}$  & 2.8$^{+2.5}_{-1.3}$ (13) & 1.3$^{+0.3}_{-0.3}$ (12) & 4.5$^{+4.5}_{-2}$ &  1.4 (23)  & $9.3 (-12)$ \\ 
Oph\,D  & 7.8$^{+0.1}_{-1.1}$  & 2.4$^{+2.1}_{-0.7}$ (13) & 2.4$^{+0.1}_{-0.7}$ (12) & 10$^{+3.5}_{-5.5}$ &  1.4 (23) & $1.7 (-11)$\\ 
\enddata 
 
\end{deluxetable}

\end{document}